\documentclass[conference]{IEEEtran}
\pdfoutput=1

\usepackage[T1]{fontenc}
\usepackage[utf8]{inputenc}

\usepackage{amsmath,amssymb,amsfonts}
\usepackage[export]{adjustbox}
\usepackage{cite}
\usepackage{graphicx}
\usepackage{placeins}
\usepackage{xcolor}
\usepackage{efbox}
\usepackage{booktabs}

\efboxsetup{linecolor=black,linewidth=0.5pt}

\usepackage{glossaries}
\glsdisablehyper
\newacronym{cots}{\mbox{COTS}}{commercial off-the-shelf}
\newacronym{dut}{\mbox{DUT}}{device under test}
\newacronym{gnss}{\mbox{GNSS}}{global navigation satellite system}
\newacronym{gps}{\mbox{GPS}}{global positioning system}
\newacronym{ocxo}{\mbox{OCXO}}{oven-controlled crystal oscillator}
\newacronym[plural=\mbox{SDRs},firstplural=software defined \mbox{radios (SDR)}]{sdr}{\mbox{SDR}}{software defined radio}
\newacronym[plural=\mbox{UAVs}]{uav}{\mbox{UAV}}{unmanned aerial vehicle}

\PassOptionsToPackage{hyphens}{url}
\usepackage{hyperref}
\hypersetup{
	pdfauthor={Julia Beuster, Carsten Andrich, Sebastian Giehl},
	pdftitle={Characterization of Lightweight GPS Disciplined Oscillators for Distributed UAV Measurement Applications},
    pdfkeywords={...}
}

\usepackage{orcidlink}

\title{
    Characterization of\\
    Lightweight GPS Disciplined Oscillators for\\
    Distributed UAV Measurement Applications
}

\IEEEoverridecommandlockouts

\author{
    \IEEEauthorblockN{%
        Julia~Beuster\raisebox{.5ex}{\orcidlink{0000-0003-1887-4278}}\IEEEauthorrefmark{1},
        Carsten~Andrich\raisebox{.5ex}{\orcidlink{0000-0002-4795-3517}}\IEEEauthorrefmark{1},
        Sebastian~Giehl\raisebox{.5ex}{\orcidlink{0009-0008-1672-1351}}\IEEEauthorrefmark{1}
    }
    \IEEEauthorblockA{\IEEEauthorrefmark{1}Institute of Information Technology, Technische Universität Ilmenau, Ilmenau, Germany}
    \thanks{This research has been partially funded by the Federal Ministry of Education and Research of Germany in the project ``6G-ICAS4Mobility'' (grant number: 16KISK241). We thank the Institute of Process Measurement and Sensor Technology at the Technische Universität Ilmenau for supporting the measurement in this publication by providing and operating the electrodynamic vibration test system. }
}

\begin{document}

\maketitle

\begin{abstract}
With an increasing variety of measurement applications using sensing nodes on \glspl{uav}, \gls{gps} and \gls{gnss} disciplined oscillators (GPSDOs, GNSSDOs) are an appealing solution for precise wireless inter-device synchronization. 
Typically evaluated under laboratory conditions by analyzing the 10\,MHz and 1~pulse per second (PPS) reference signal stability, these test methods overlook airborne oscillator performance.
This paper characterizes lightweight \mbox{GNSSDO} models for flight suitability using a measurement system based on \glspl{sdr}. 
We analyze reference signal stability under controlled \gls{gnss} signal impairments to predict performance and measurement precision loss in dynamic operational modes. 
Additionally, we assess behavior under the impacts caused by operating the UAV, as well as typical flight vibrations and accelerations outlined in the standard for payload devices.
\end{abstract}


\section{Introduction}
The growing deployment of \glspl{uav} in various measurement applications has emphasized the need for precise timing and synchronization mechanisms, often provided by small, lightweight GNSSDOs. Despite their importance, the performance of GNSSDOs under the dynamic conditions of UAV flights remains inadequately characterized in existing literature.

To address this gap, we introduce a comprehensive laboratory GNSSDO testbed designed to analyze the stability of 10\,MHz and 1\,PPS reference signals specifically for UAV applications. 
We exemplify its capabilities by characterizing the behavior of two \gls{ocxo}-based GNSSDO models for short set-up times, in steady state, during \gls{gnss} signal impairments, under mechanical disturbances, and under the influence of UAV electronics.

\section{GNSSDO testbed}
To comprehensively assess and enhance the performance of \mbox{GNSSDOs} for the utilization in UAV applications, it is essential to precisely analyze the time error between 10\,MHz and 1\,PPS reference signals concurrently among multiple oscillators, while considering \gls{gnss} signal quality and the distinct effects encountered during flight.
We realized this by extending our GNSSDO testbed~\cite{ifcs20bauerlab} with an electrodynamic vibration test system and a quadcopter mock-up, as \mbox{shown in \autoref{testbed}}.

\subsubsection{Time and frequency stability measurement system}
Our measurement system supports simultaneous analysis of 1\,PPS and 10\,MHz reference signal stability using only \gls{cots} components and digital signal processing (DSP) software~\cite{8288589}.
Its most recent hardware iteration comprises two \textit{USRP X310} from Ettus \mbox{Research\texttrademark} and enables to characterize the performance of up to four \mbox{GNSSDOs} with sub-nanosecond precision~\cite{ifcs20bauerlab}.

\subsubsection{Electrodynamic vibration test system}
The vibration test system is based on the electromechanical \textit{Shaker V300} from Data \mbox{Physics\texttrademark \cite{lang2001understanding}}, which allows to accelerate payload with up to 98g and a maximum sine force of 1646\,N in a frequency range from 0 to 5000\,Hz and a shock force of up to 3490\,N.

\subsubsection{Quadcopter mock-up}
To emulate the impact of UAV electronics on \mbox{GNSSDOs}, a fully operational and remotely controllable drone mock-up with a cabled power supply is needed to expose the \gls{dut} to reproducible and repeatable conditions, enabling automated long-term measurement cycles.
The presented mock-up is built upon a \textit{Tarot Iron Man FY650} frame with a diameter of 650\,mm using the propulsion system \textit{DJI E800}, making it comparable in size and motors to DJI's 3--4\,kg camera drones, such as the \textit{DJI Inspire I}. 
Customized control electronics enable reliable thrust settings provided to the electronic speed controllers of individual rotors by periodic pulses with widths between 1100\,\textmu{}s at 0$\%$ and 2000\,\textmu{}s at 100$\%$~\cite{mockup}.

\begin{figure}
\centering
\includegraphics [width=\linewidth] {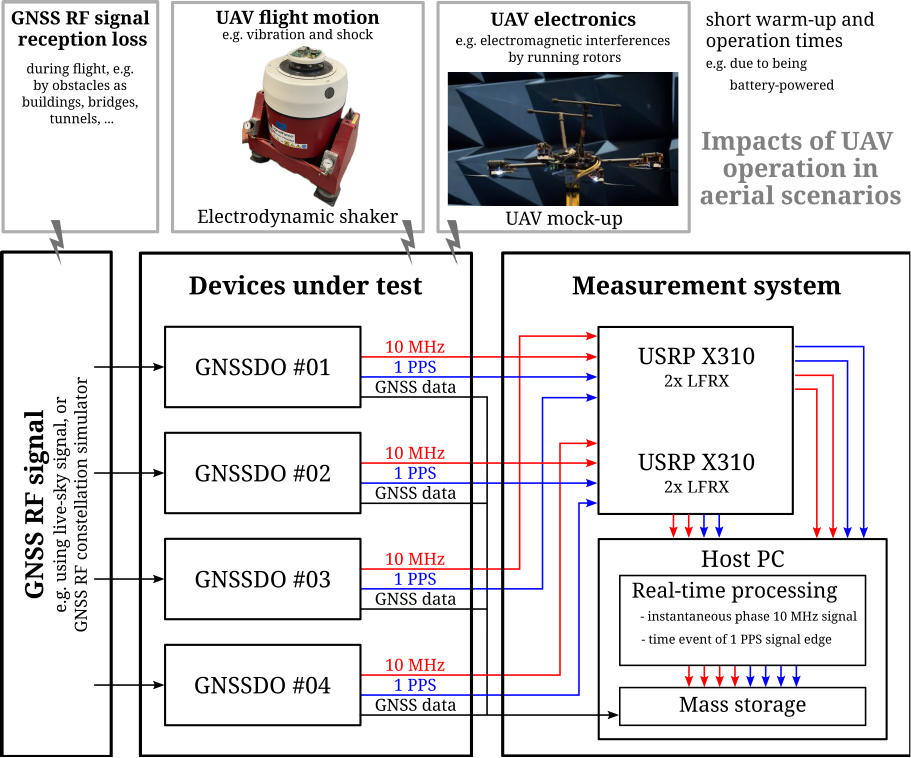}
\caption{Simplified block diagram of a laboratory \mbox{GNSSDO} measurement testbed that enables the simultaneous analysis of the reference signal stability of 10\,MHz and 1\,PPS signals of up to four \mbox{GNSSDOs} under test with sub-nanosecond precision for long-term observation intervals using a DSP- and SDR-based measurement setup~\cite{ifcs20bauerlab, 8288589} in combination with an electrodynamic vibration test system, a so called shaker, and a quadcopter mock-up to expose the \glspl{dut} to conditions present when operating UAVs in aerial scenarios.}
\label{testbed}
\end{figure}

\section{Evaluation of GNSSDO performance}
To showcase the functionality of the introduced \mbox{GNSSDO} testbed, two devices from each of two widely-used, low-cost, lightweight, OCXO-based \mbox{GNSSDO} models available to the authors are characterized in terms of their reference signal stability in different operation modes.
These \mbox{GNSSDO} models, namely the \textit{Jackson Labs GPSDO LC$\_$XO OCXO}~\cite{manualLCXO} and the \textit{Furuno Multi-GNSSDO GF-8805}~\cite{manualGF8805}, were selected because of their promising values in terms of weight, spatial dimensions, power consumption, and reference signal stability specified in the data sheets.
For the presented measurement data, all \glspl{dut} are subjected to the same live-sky \gls{gnss} scenario using a stationary, exposed antenna position with clear sky view on the rooftop of a laboratory building.

\subsection{Reference signal stability in steady-state}
\begin{figure}
\centering
\includegraphics [width=0.85\linewidth] {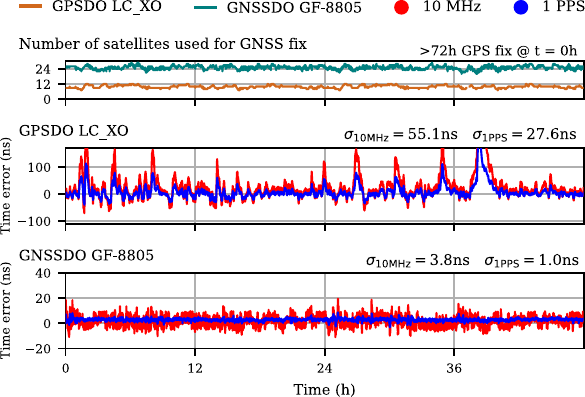}
\caption{Steady-state timing accuracy provided by lightweight \mbox{GNSSDO} models available to the authors in context of reception quality in a stationary, live-sky \gls{gnss} signal scenario. The time error between the 10\,MHz and 1\,PPS reference signals is determined between two \glspl{dut} of each type (see \autoref{testbed}, \mbox{GNSSDO}~$\#$01..02: \textit{Jackson Labs GPSDO LC$\_$XO OCXO}, GNSSDO~$\#$03..04: \textit{Furuno GF-8805}). All devices had a valid \gls{gnss} fix for more than 72 hours at the start of the measurement.}
\label{steadystate1}
\end {figure}

\begin{figure}
\centering
\includegraphics [width=0.9\linewidth] {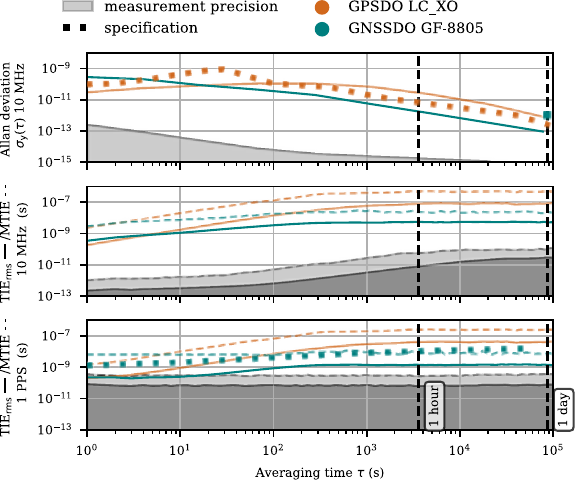}
\caption{Comparison of measured 10\,MHz and 1\,PPS reference signal stability of \mbox{GNSSDOs} under test in steady-state with manufacturer specifications in terms of Allan deviation, TIE$_\mathrm{rms}$, and MTIE.}
\label{steadystate2}
\end {figure}

In order to enable the selection of \mbox{GNSSDO} devices from the multitude of models on the market, the reference signal stability of oscillators in steady-state under laboratory conditions is specified in data sheets and academic research 
using the statistical time-domain metrics Allan deviation $\sigma_\mathrm{y}(\tau)$ and time interval error (TIE) in form of its maximum (MTIE) and root-mean-square (RMS) value TIE$_\mathrm{rms}$.
As presented in \autoref{steadystate1}, after being locked for more than 72\,hours to the timing information derived from the \gls{gnss} signals, the time errors of the \mbox{GNSSDO} models under test show clear differences within an order of magnitude.
The time error between the 1\,PPS reference signals provided by the GPSDO model \textit{LC$\_$XO} is strongly affected by outliers and varies with a standard deviation $\sigma$ of up to~28\,ns, whereas the \textit{GF-8805} provides 1\,PPS signals with a time error $\sigma$ of only up to~1\,ns.
As shown in \autoref{steadystate2}, the short-term stability of the 10\,MHz and 1\,PPS reference signals provided by the \mbox{GNSSDO} \textit{GF-8805} is worse than that of the GPSDO \textit{LC$\_$XO}, but outperforms the \textit{LC$\_$XO} for long-term observation intervals.
The maximum time error of both reference signal types provided by the \textit{LC$\_$XO} exceeds 100\,ns over a time interval $\tau$ of 30\,minutes, which matches a common flight duration, whereas for the \textit{GF-8805} the corresponding MTIE is only 10\,ns for the provided 1\,PPS reference signals and 25\,ns for the 10\,MHz reference signals.

\subsection{GNSSDO performance for short set-up times}
\begin{figure*}[h]
\includegraphics [width=0.9\linewidth] {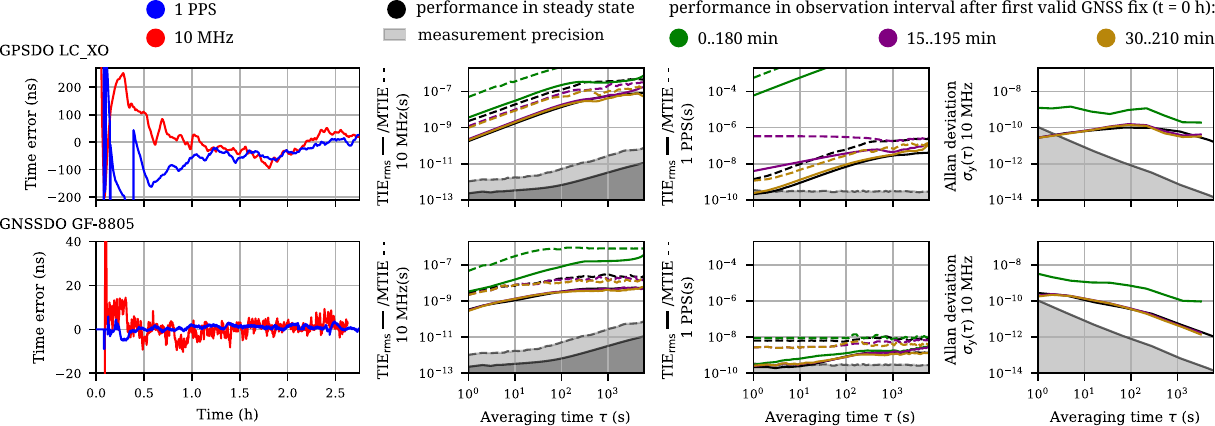}
\caption{Time error and reference signal stability in terms of Allan deviation, TIE$_\mathrm{rms}$, and MTIE of \glspl{dut} for different observation intervals after being cold started. It is apparent that the \mbox{GNSSDOs} are strongly affected by oscillator settlement in the initial warm-up phase and at least 15\,min for the \textit{GF-8805} or 30\,min for the \textit{LC$\_$XO} after the first valid \gls{gnss} fix are required to ensure that the \glspl{dut} deliver comparably stable 10\,MHz and 1\,PPS reference signals.}
\label{warmup}
\end {figure*}

In addition to the performance during steady-state operation, certain measurement applications, particularly those utilizing battery-powered systems as it is common in UAV scenarios, necessitate the \mbox{GNSSDOs} to be set up shortly before usage, thus requiring examination of their performance in the stabilization interval following a cold start.
As depicted in \autoref{warmup}, the 10\,MHz and 1\,PPS reference signals provided by both \mbox{GNSSDO} models under test exhibit noticeable settling drift.
To ascertain the duration between the initial valid \gls{gnss} fix and the effective usability for a specific measurement use case, it is necessary to evaluate the influence of oscillator settling on the resultant time error.
Comparing the reference signal stability of the \glspl{dut} in steady-state with the stability of devices affected by settling behavior, as shown in \autoref{warmup}, in terms of Allan deviation and TIE, it becomes apparent that the \textit{GF-8805} requires up to 15\,min to stabilize. The \textit{LC$\_$XO} needs at least 30\,min to stabilize and provide acceptable reference signals.

\subsection{Oscillator behavior during GNSS signal impairment}
\begin{figure}
\centering
\includegraphics [width=0.85\linewidth] {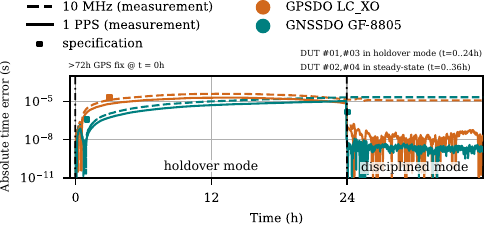}
\caption{Time error between 10\,MHz and 1\,PPS reference signals to characterize the behavior of the oscillators models operating in holdover mode and the transition behavior when returning to \gls{gnss} disciplined mode. \gls{dut} $\#$01 and $\#$03 are affected by a 24\,hour long \gls{gnss} signal reception loss, whereas \gls{dut} $\#$02 and $\#$04 are operated in steady-state. All devices had a valid \gls{gnss} fix for more than 72\,hours at the start of the measurement.}
\label{holdover}
\end{figure}

\gls{gnss} signal impairments, e.g., line-of-sight obstruction by buildings, bridges, or vegetation, are inherent to non-stationary \mbox{GNSSDO} applications.
Typical devices provide only limited information on their output reference signal quality or require to infer the output signals' accuracy from \gls{gnss} signal quality metrics.
To select suitable \mbox{GNSSDOs} based on performance with impaired \gls{gnss} signal, both holdover behavior and the return to disciplined mode must be analyzed.
The resulting time error during the absence of \gls{gnss} signals depends mainly on the \mbox{GNSSDO's} implementation to bridge the reference signals until the next \gls{gnss} fix, as shown in \autoref{holdover} and \autoref{tab:GPSDO_holdover}.
For both OCXO-based \mbox{GNSSDO} models under test, the time error of the 10\,MHz and 1\,PPS reference signals increases strongly during a \gls{gnss} signal loss. 
After being re-disciplined the 1\,PPS reference signals of both models recover the timing accuracy before the \gls{gnss} signal outage, while the drift of the 10\,MHz signals accumulated during the holdover operation results in a significant time error offset. 

\begin{table} [h]
    \caption{Holdover stability 10\,MHz reference signal} \vspace{-1em}
    \centering
    Maximum time error ($\mu$s) \vspace{0.5em} \\
    \scriptsize
    \begin{tabular}[width=\linewidth]{lcccccc}
        \toprule
        GNSSDO & @ 15min & @ 30min & @ 1h & @ 3h & @ 12h & @ 24h \\
        \midrule
        \textit{LC$\_$XO} & 0.2 & 0.3 & 2.9 & 13.4 & 40.9 & 41.3 \\
        \textit{GF-8805} & 0.1 & 0.2 & 0.2 & 1.1 & 10.4 & 21.9\\
        \bottomrule
    \end{tabular}
    \label{tab:GPSDO_holdover}
    \end{table}

\subsection{GNSSDO performance under mechanical disturbances}
\begin{figure}
\centering
\includegraphics [width=\linewidth] {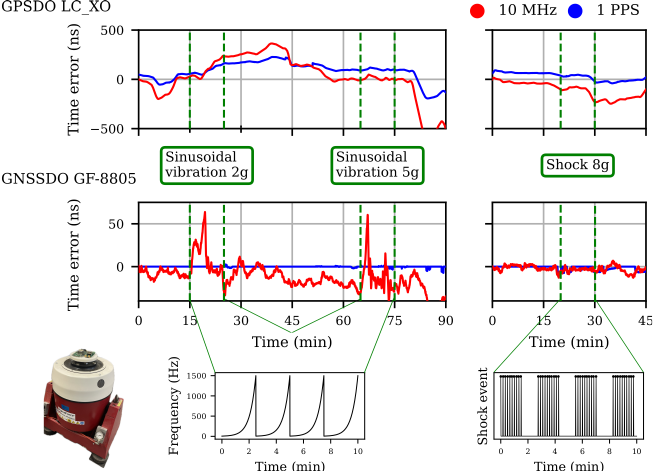}
\caption{Time error between 10\,MHz and 1\,PPS reference signals of one device per \mbox{GNSSDO} type (\gls{dut} $\#$02 and $\#$04) in steady-state and one device per model (\gls{dut} $\#$01 and $\#$03) influenced by vibration and shock parameters outlined in standard for UAV payload devices~\cite{9354136}. The operational sequences of vibration and shock were exemplarily chosen to repeat four times over a time period of 10\,min. Note that for the \mbox{GNSSDO} model \textit{GF-8805} the impact of the sinusoidal vibrations results in outliers during the exposure, whereas the \textit{LC$\_$XO} shows strong settlement behavior even after the end of the exposure.}
\label{shaker}
\end {figure}

To mitigate the impact of flight motions on \gls{uav}-mounted devices, countermeasures are necessary. To reduce development effort and costs, while considering UAV payload weight and space constraints, examining these mechanical effects on \mbox{GNSSDO} model performance is beneficial.
\autoref{shaker} exemplifies the time error between reference signals of one \mbox{GNSSDO} per model in steady-state and one \gls{dut} per model influenced by conditions of UAV operation in form of vibration and shock parameters outlined in standard requirements for UAV payload devices~\cite{9354136}.
It is apparent that the reference signal stability of both \mbox{GNSSDO} types is affected by the sinusoidal vibrations during the exposure. Additionally, the GPSDO \textit{LC$\_$XO} shows strong drifting behavior for both reference signals after the end of the mechanical exposure.

\subsection{Reference signal stability during UAV mock-up operation}
\begin{figure}
\centering
\efbox{\includegraphics[width=0.85\linewidth] {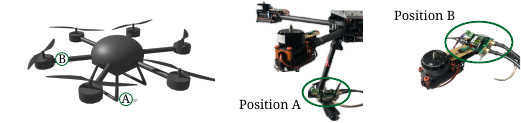}} 

\vspace{0.5em}

\includegraphics[width=0.9\linewidth] {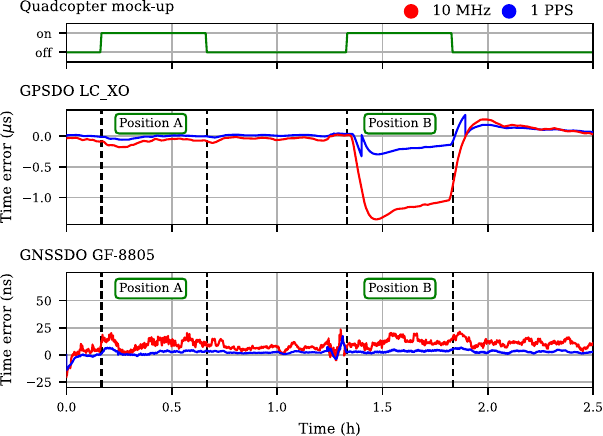}
\caption{Time error between reference signals of one device per \mbox{GNSSDO} type (\gls{dut}~$\#$02 and $\#$04) in steady-state and one (\gls{dut}~$\#$01 and $\#$03) influenced by operating a \gls{uav} mock-up. The operational sequences of alternating thrust values were exemplarily chosen to resemble a 30\,min long flight operation. Note that for the \mbox{GNSSDO} model \textit{GF-8805} the impact of the UAV electronics results in no outliers during the exposure, whereas the \textit{LC$\_$XO} shows outliers and settlement behavior corresponding to the distance to the rotors.}
\label{mockup}
\end {figure}

Due to strict space and weight limits for UAV payload, determining device placement and the necessity for additional electromagnetic shielding are crucial considerations when planning applications with participating UAV nodes.
The quadcopter mock-up shown in \autoref{testbed} can be used to exemplify the impact of the UAV electronics on the reference signal stability of the \mbox{GNSSDOs} under test for different positions during flight. 
Two extreme placement options were tested: Position~(A) is located at the bottom of the potential payload space to maximize the distance between the \glspl{dut} and the running rotors with their power supply cables, whereas position~(B) is chosen to place the \glspl{dut} as close to the rotors as possible.
The motor thrust values provided to the electronic speed controller of the UAV mock-up were chosen to alternate every 20\,seconds between periodic pulse widths of 1500\,\textmu{}s and 1600\,\textmu{}s to emulate a real flight scenario with straight motion vectors interrupted by changes in direction.
The time error between the reference signals of one \mbox{GNSSDO} under test in
steady-state and one influenced by conditions of the operating UAV, as shown in \autoref{mockup}, reveals clear differences in the robustness of the two \mbox{GNSSDO} model types.
The \mbox{GNSSDO} model \textit{GF-8805} shows no significant outliers during the exposure, whereas the GPSDO \textit{LC$\_$XO} shows outliers and settlement behavior after the end of the exposure corresponding to its distance to the rotors.

\section{Conclusion}
This paper introduced a laboratory testbed to analyze the 10\,MHz and 1\,PPS reference signal stability of \mbox{GNSSDOs} for UAV applications.
The testbed combines a highly precise SDR- and DSP-based reference signal stability measurement system, an electrodynamic vibration test system, and a UAV mock-up to characterize the performance of up to four \glspl{dut} simultaneously. 
This approach isolates and enables the analysis of the impact of individual effects occurring during UAV flight operations, allows to improve the reference signal stability by targeted countermeasures, and complements specification sheets to aid in the selection of appropriate devices for synchronizing mobile aerial nodes.
To demonstrate the testbed’s capabilities, we characterized the reference signal stability of two low-cost, lightweight, OCXO-based \mbox{GNSSDO} model types.
This characterization included behavior relative to \gls{gnss} signal reception quality in steady-state, warm-up, and holdover mode, as well as conditions of UAV operation.
The more affordable of the two \mbox{GNSSDO} models exhibited better signal stability and robustness. However, its larger size and weight limits its suitability for small \glspl{uav}. 

\bibliographystyle{IEEEtran}
\bibliography{paper}

\end{document}